\documentclass[11pt]{article}
\usepackage{graphicx}
\author{Kameshwar C. Wali\\
 Physics Department,syracuse university, Syracuse, New York}

\title{Clash of Symmetries in a Brane World Picture\footnote{Paper presented at the
IV International Symposium, Quantum Theory and Symmetries, 15-21
August 2005, Varna, Bulgaria.}}

\begin{document}
\maketitle
 \large

 \begin{abstract}

If our (3+1) dimensional universe is a brane or domain wall
embedded in a higher dimensional space, then a phenomenon that may
be designated as "Clash of Symmetries" provides a new method of
breaking continuous symmetries. The paper presents some
non-trivial models containing the physical ideas.

 \end{abstract}

\section{Introductory remarks}

Symmetry, as wide or as narrow as you may define its meaning, is
one idea by which man through the ages has tried to comprehend and
create order, beauty, and perfection

\begin{flushright}  \emph{Herman Weyl} in Symmetry
\end{flushright}
\vspace{0.5cm}

We may hope that the remaining two decades of our century, which
began under the sign of symmetry, physicists will be able to
explain how great variety of non-symmetrical forms of the real
world can arise out of the beautiful,symmetrical structure of the
basic equations. At the beginning of the century, Einstein taught
us to understand the meaning of symmetry; we have now to learn how
to break it in the most symmetrical way so that his most cherished
dream, that of building a unified theory of all interactions, may
one day come true.

\begin{flushright}        \emph{Luigi Radicati di Brozol}
\end{flushright}
\vspace{0.5cm}

These two statements express two profound and basic aspects of
physics today. Weyl, in his beautiful and elegant book, described
as his swan song, elaborates on the role of symmetries in Nature,
in art and architecture, and then shows how they are transformed
in mathematical forms to express laws of physics. Radicatti, on
the other hand, stresses the equally important fact of Nature,
namely, the real world does not exist in the perceived perfect
symmetric form, but more naturally in its "approximate" or"broken"
forms. While we can build beautiful models of elementary particles
and their interactions based on exact symmetries, real world
defies them and remains provocative. Whether it be the Standard
Model and beyond, SUSY, String theory or any other theory,
symmetry breaking poses as a fundamental problem.

In what follows, I will briefly and qualitatively describe a new
approach to symmetry breaking that has been the subject of study
during the past few years \cite{Davidsonet al}. What I am going to
describe are some non-trivial, but still in the nature of toy
models. We, think, however, they do indicate a new and fruitful
approach to realistic models

\section{Clash of symmetries; basic idea}

The conventional spontaneous symmetry breaking mechanism (SSB)in
the standard model(SM) and its extensions assumes spatially
homogeneous Higgs fields whose vacuum expectation values are
determined by the minimization of a postulated Higgs potential.
The non-vanishing constant values determine the masses of the
gauge and matter fields through their couplings to the Higgs
fields. The considerable arbitrariness involved in the choice of
the potentials and the profusion of free parameters make this
scheme unsatisfactory.

The non-commutative geometric framework of Alain Connes has
provided a new and elegant scheme that elevates SSB to a new level
of its understanding. This is accomplished by placing Higgs fields
and gauge fields on a similar geometrical footing. This makes the
models more predictive. However, this framework beyond its
application to SM, requires departures from from the original
rigorous approach and poses some problems that I cannot go into in
this brief report.

There are, however, other types of solutions to Higgs fields that
can serve as stable, static background fields: topological
solitons, such as kinks, strings and monoploes. But their
spatially non-homogeneous nature forbids them to be used as
background fields in our 3+1 dimensional space-time universe,
since they conflict with the strong evidence of large sale
homogeneity. This objection, however, does not apply to
brane-world models, since the non-trivial spatial dependence of
the Higgs fields can be restricted to extra dimension co-ordinates
only.

Consider, for instance, an extra dimension, coordinate $w$ and
topologically stable Higgs field configurations $\phi_{i}(w)$,
some of which have kink form with respect to $w$. Then the pattern
of symmetry breaking becomes a function of $w$. Suppose that the
3+1 dimensional brane world is located at $w=0$ with a set of
localized physical fields confined to the brane at $w=0$. If the
fields are strictly confined to $w=0$, then the unbroken symmetry
is the stability group of $\phi_{i}(w=0)$, say a subgroup $H(w=0)$
of some internal symmetry group $G$. However, in quantal(or
perhaps even in classical world), one would not expect the fields
absolutely confined, in which case they couple , perhaps with
reduced strength, to $\phi_{i}(w)$ states, where $w$ is different
from $H(w=0)$.If the stability group
$\phi_{i}(0<|w|<\epsilon),h(w\sim\epsilon)$ is different from
$h(w=0)$, then a rich effective symmetry breaking is possible on
the brane

This happens when isomorphic groups $H(|w|=\infty$) left unbroken
at $|w|=\infty$ \emph{can be differently embedded in the parent
group $G$}. The break down at finite $w$ is the intersection of
the asymptotic stability groups
$$
H(|w|<\infty)=H(-\infty)\cap H(+\infty)\equiv H_{clash}
$$

Although $H(+\infty)$ and $H(-\infty)$ are isomorphic, their
different embeddings leads to their intersection that is a smaller
group.

Thus in general, if the vacuum manifold is disconnected and
contains distinct vacuum states, the field can settle into
different minimizing vacuum states in different spatial
dimensions. Stable solitonic or kink-like domain wall
configurations can exist if the vacuum manifold has the
appropriate topology.

If our (3+1) dimensional universe is a brane or a domain wall,
such a situation can arise, namely, a global $G_{cts}\otimes
G_{discrete}$ is spontaneously broken
 to  $H_{cts}\otimes H_{discrete}$,
where $H_{cts}$ can be embedded in several different ways in the
parent $G_{cts}$and {$H_{discrete} \subset G_{discrete}$}.

As a consequence, a certain class of domain wall solutions
connects two vacua that are invariant under differently embedded
$H_{cts}$ subgroups and there is an enhanced symmetry breakdown to
the intersection of the two subgroups on the brane or domain wall.
 In the brane limit, {$H_{cts}$} prevails in the bulk, but smaller
intersection symmetry on the brane itself.

One may call this phenomenon \emph{CLASH OF SYMMETRIES}.

\section{A model with three higgs triplets}

Consider a model with three Higgs triplets $\phi_{1,2,3}$
interacting through the potential
$$
 {V}= {-\sum_{i=1}^{3} (\phi^{\dagger}_{i}\phi_{i})+ (\sum_{i=1}^{3}
\phi^{\dagger}_{i}\phi_{i})^2 + V_{4}}
$$
$$
{V_{4} }={\lambda/2\sum_{i\not=j}
(\phi^{\dagger}_{i}\phi_{i}\phi^{\dagger}_{j}\phi_{j})
 +\sigma/2\sum_{i\not=j}
(\phi^{\dagger}_{i}\phi_{j}\phi^{\dagger}_{j}\phi_{i})}
$$
 The symmetry group of the potential is
 $ G=G_{cts}\otimes G_{discrete}$, where
$ G_{cts}=SU(3)\otimes U(1)_{1}\otimes U(1)_{2}\otimes U(1)_{3}$
and {$G_{discrete}=S_{3}$}.

 Minimization of the potential yields three
degenerate vacua, leading to a manifold consisting of three
disconnected pieces,
$$
vacuum i: (\phi^{\dagger}_{1}\phi_{1})=1/2,
(\phi^{\dagger}_{2}\phi_{2})=(\phi^{\dagger}_{3}\phi_{3})=0
$$
with the other two obtained by permutations.

Each global minimum induces the SSB of both the parent discrete
and continuous symmetries to {$ S_{3}\rightarrow S_{2}$},
{$G_{cts}\rightarrow U(2)\otimes U(1)^{2}$}.

\begin{figure}[t!] 
\centerline{
   \includegraphics[height=2.in, width=2.in ]{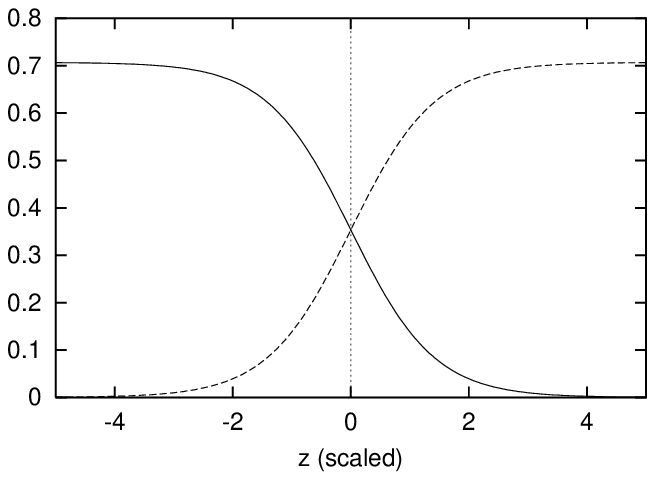}
   \hspace{2.cm}
   \includegraphics[height=2.in, width=2.in ]{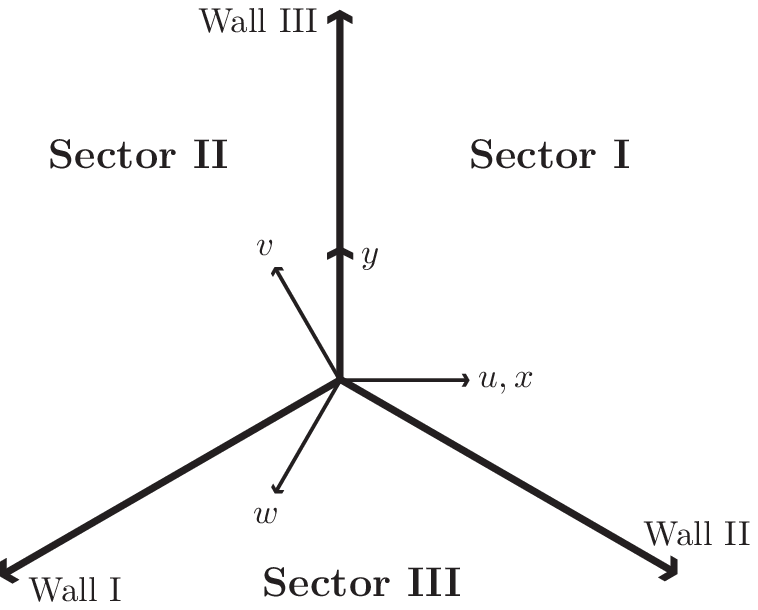}
   }
 \caption {Left: Typical kink-type solution. Right: The three--star domain wall
 junction configuration.}
\label{kinksol}
\end{figure}

A kink or one-dimensional domain wall configuration interpolates
between the elements of disconnected vacua. A typical solution is
shown in Fig. \ref{kinksol} (left panel).

 A more interesting domain wall junction configuration is depicted
 Fig \ref{kinksol} (right panel).
  Three semi-infinite walls meet at a point, the origin or
 nexus, at angles of $2\pi/3$. Ignoring the superfluous $U(1)$'s,
 the clash of symmetries has the pattern:

 $$
  H_{I\cap II}=U(2)_{II}\cap U(2)_{II}=U(1)_{III}
 $$
 along wall $III$ with corresponding results for $H_{II\cap III}$
 and $H_{III\cap I}$

\begin{figure}[b!] 
\centerline{
   \includegraphics[height=2.in, width=2.in ]{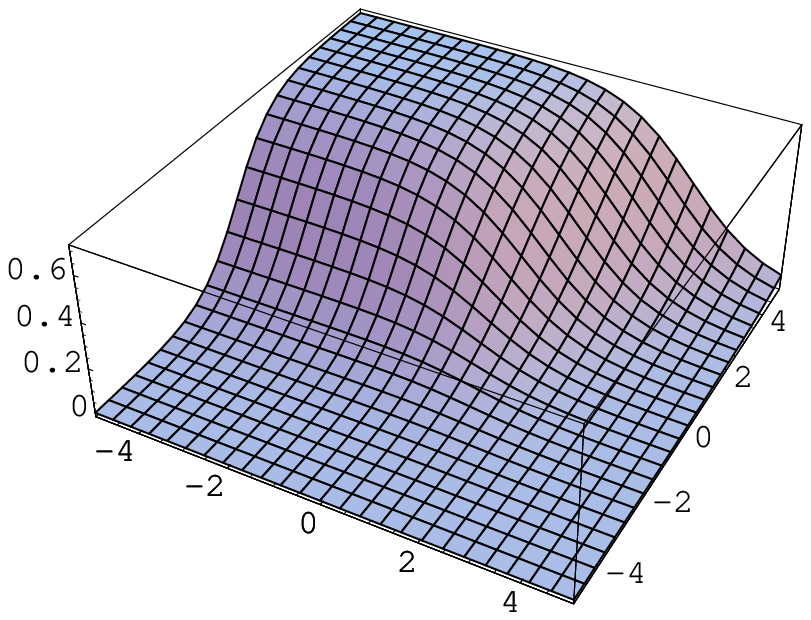}
   \hspace{.5cm}
   \includegraphics[height=2.in, width=2.in ]{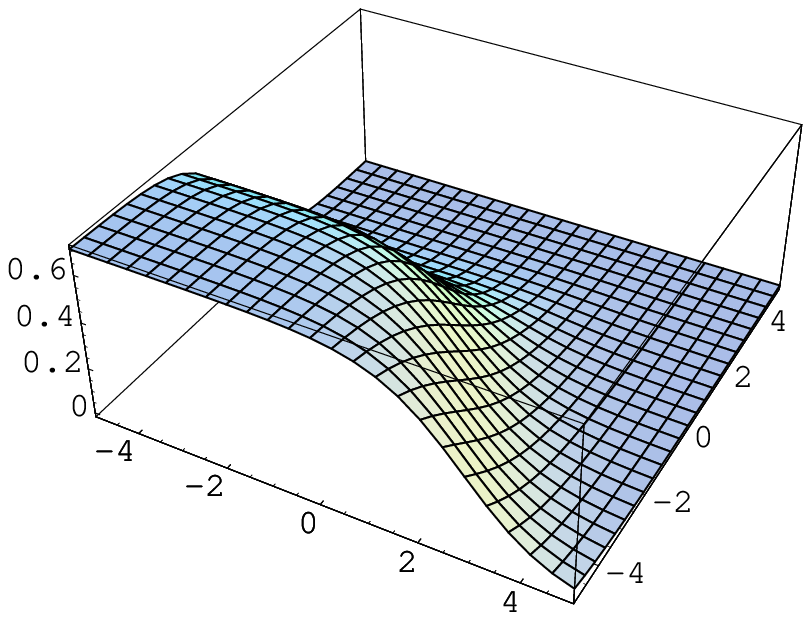}
  \hspace{.5cm}
   \includegraphics[height=2.in, width=2.in ]{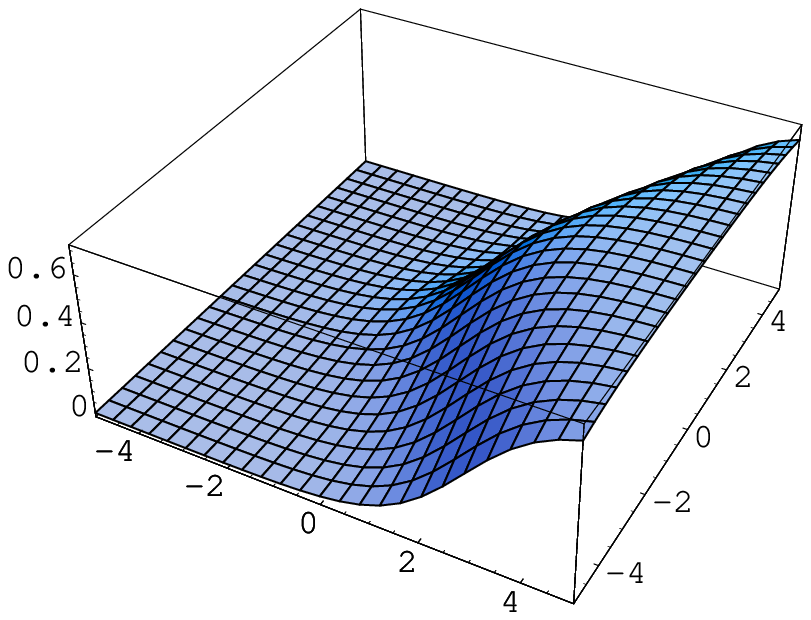}
   }
 \caption {$\phi_1, \phi_2, \phi_3$ components of the three--star configuration.}
\label{gmin}
\end{figure}

Figures of global minima are shown in Figs. \ref{gmin}.

\section{Domain wall solutions with Abelian gauge fields}

The lagrangian consists of two complex scalar fields $\phi_{1.2}$
coupled to Abelian gauge fields $a_{1\mu}, a_{2\mu}$
$$
{\cal{L}}= -(1/4)(\sum (F_{i}^{\mu\nu}F_{i\mu\nu})+
\sum(D^\mu\phi^{\dagger}_{i})D_\mu\phi_{i})-V(\phi_{1},\phi_{2})\\
$$
$$
V(\phi_1,\phi_2)=
\lambda_{1}(\phi^{\dagger}_{1}\phi_{1}+\phi^{\dagger}_{2}\phi_{2}-v^2)^2
+\lambda_{2}(\phi^{\dagger}_{1}\phi_{1}\phi^{\dagger}_{2}\phi_{2})
$$
with the overall continuous symmetries {$u(1)\otimes u(1)$} and
the discrete symmetry {$ z_{2}$}.

\begin{figure}[h!] 
\centerline{
   \includegraphics[height=2.in, width=2.in ]{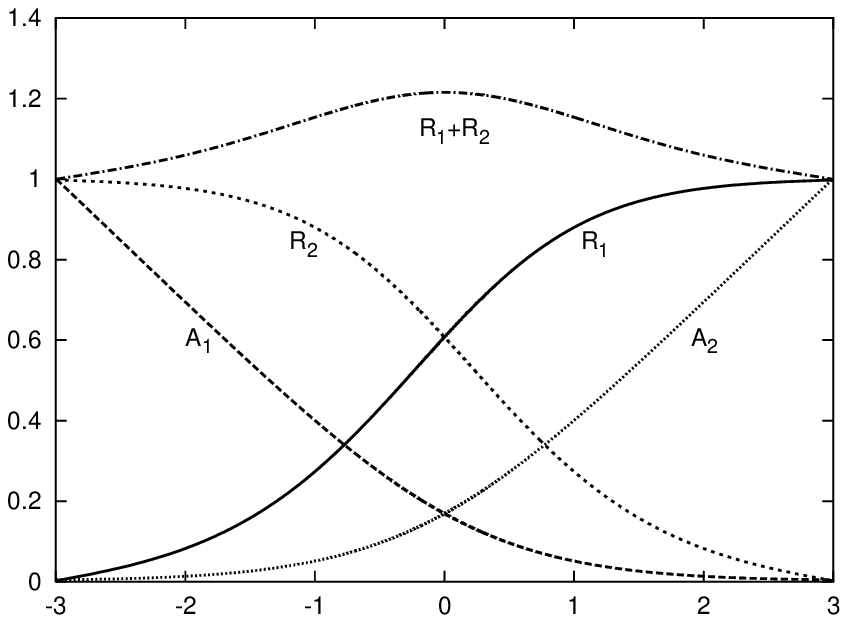}
   \hspace{2.cm}
   \includegraphics[height=2.in, width=2.in ]{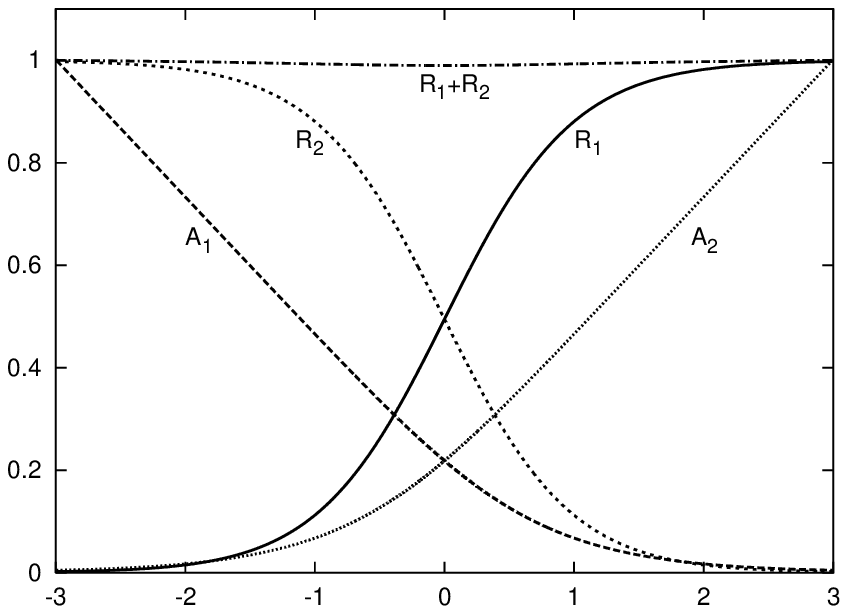}
   }
 \caption {Plots of scalar and gauge fields for different sets of
 parameters}
\label{Aplot}
\end{figure}

There are two distinct vacua. With appropriate boundary conditions
for the scalar fields and the implied boundary conditions for the
gauge fields, we find numerical solutions for the coupled
equations. We obtain expected kink solutions for the scalar
fields. The gauge fields diverge linearly on either side, but fall
off exponentially on opposite sides (Fig. \ref{Aplot}). The
$U(1)$symmetries are preserved in their respective vacua but
broken elsewhere. The domain wall is sandwiched between domains of
constant magnetic fields parallel to the wall. In the case of a
domain wall with finite thickness, there will be magnetic fields
parallel to the wall on either side associated with
superconducting currents, as in the case of the superconducting
string solutio

Thus, in addition to symmetry breaking on the brane, we find a new
phenomenon such as the appearance of magnetic fields in the bulk.

\section{A global $U(1)\otimes U(1)$ model with
Randall-Sundrum-like gravity}

 The starting point is a model with two complex scalar fields in a five dimensional
space-time with the action:
$$
{\cal{S}} = \int {[-{\kappa\cal {R}}/2 \ -\ {\cal
L}(\phi_{1},\phi_{2})] \sqrt{-g} } \ d^4x \ dw,
$$
where
$$
{\cal L}(\phi_{1},\phi_{2}) =
{g^{ab}\sum(D_{a}\phi_{i})^*(D_{b}\phi_{i})}+V
$$
and the metric is given by
$$
{ds^2} = {dw^2 + e^{2f(w)} \eta^{\mu\nu}dx_{\mu}dx_{\nu}}
$$
In addition to the $U(1)\otimes U(1)$ symmetry, there is a
discrete symmetry $\phi_{1}\leftrightarrow \phi_{2}$. The field
equations that follow are
$$
{2V(\phi_{1},\phi_{2})}= {-3\kappa(f^{''}+4(f^{'})^2)}
$$
$$
{\sum(\phi_{i}^*\phi_{i}^{'})}={-3\kappa f^{''}}
$$
$$
{\phi_{i}^{''}+4f^{'}\phi_{i}^{'}} = {(\delta
V)/(\delta\phi_{i})},
$$
where 'prime ' denotes differentiation with respect to $w$.

We seek static solutions of the field equations with $w$ as the
variable or alternately we attempt to find a potential that
satisfies the equations. With the ansatz
$$
{\phi_{1}}={(v/\sqrt(2))\sqrt(1+\tanh\beta w)}
$$
$$
{\phi_{2}}={(v/\sqrt(2))\sqrt(1-\tanh\beta w)}
$$
 we obtain an analytic solution
$$
{e^{2f(w)}}={(\cosh\beta w)^{-(\beta v^{2})/(6\kappa)}},
$$
which has the Randall-Sundrum limit
$$
{e^{2f(w)}}\rightarrow {e^{-((v^2\beta)/(6\kappa))|w|}}
$$
provided,
$$
{{\beta}\rightarrow {\infty}}, v \rightarrow {0},
{v^2\beta}\rightarrow {finite}\ .
$$

The sextic potential that satisfies all the field equations and is
both bounded from below and has the desired global minima:
$$
{V} = -{{\beta^2v^4}\over
{24\kappa}}+{{\beta^2}\over{2v^2}}(1+{{v^2}\over{3\kappa}})\phi_{1}^{2}\phi_{2}^{2}+U+W,
$$
where
$$
{U}=
-{{\beta^2}\over{v^2}}({{3}\over{2}}+{{v^2}\over{3\kappa}})\phi_{1}^{2}\phi_{2}^{2}(\phi_{1}^{2}+\phi_{2}^{2}-v^2)
$$
and
$$
{W}= \zeta{{\beta^2\over{4v^2}}}({{3}\over{2}}+{{v^2}\over
{3\kappa}})(\phi_{1}^{2}+\phi_{2}^{2}-v^2)^2
(\eta+{{\phi_{1}^{2}+\phi_{2}^{2}-v^2}\over{v^2}}),
$$
 $\zeta$ and $\eta$ are parameters such that
$$
\zeta\geq 1; \    \zeta\eta > 1 \ .
$$

The equations are satisfied without $W$, but then the potential is
not bounded from below. We note that there is a negative
cosmological constant in the bulk
$$
{\lambda_{5}} = {-{{\beta^2v^4}\over{24\kappa}}} \ .
$$

\begin{figure}[h!] 
\centerline{
   \includegraphics[height=2.in, width=2.in ]{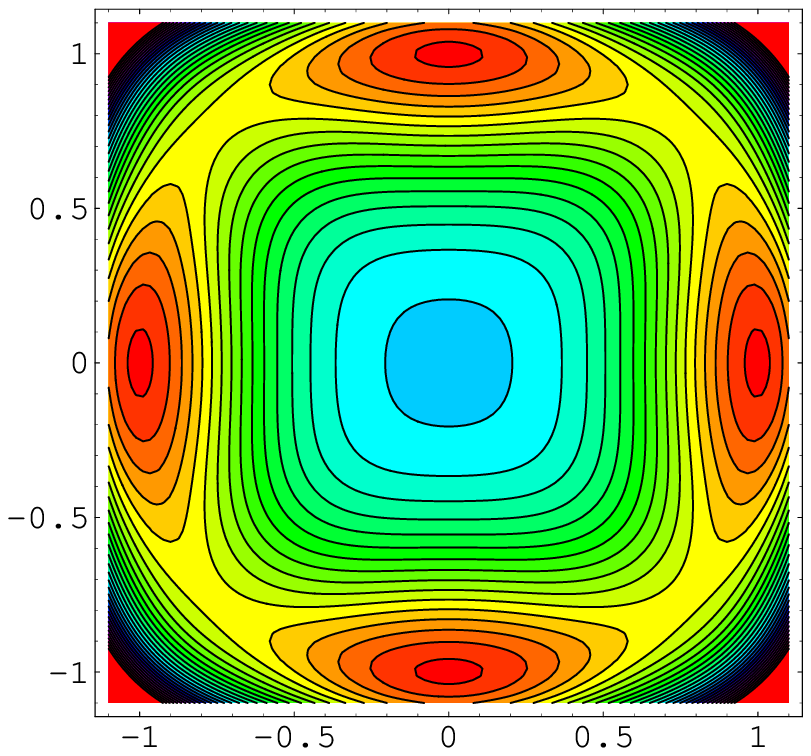}
   \hspace{2.cm}
   \includegraphics[height=2.in, width=2.in ]{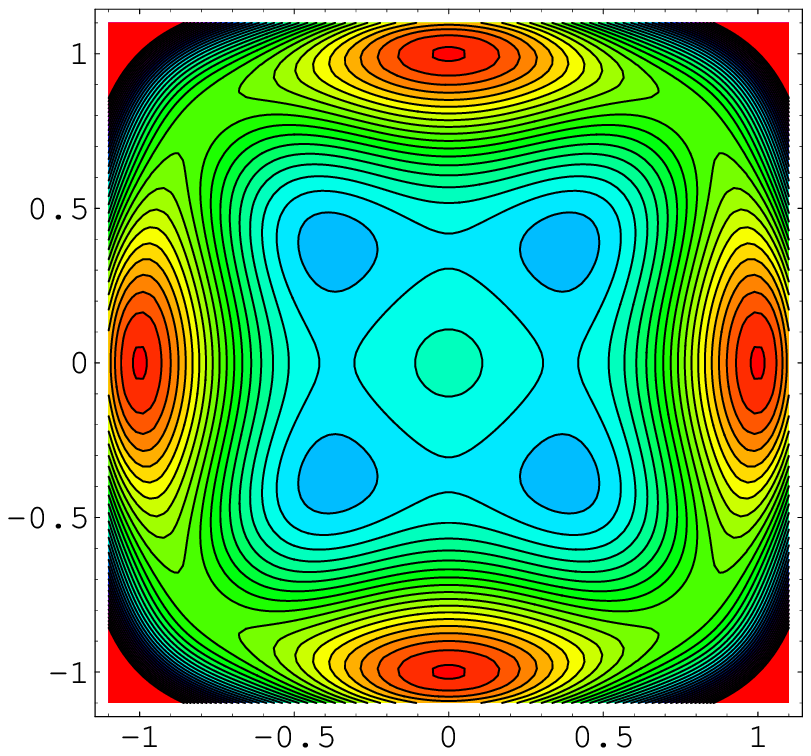}
   }
 \caption{Contours of the qualitative features of the extrema of
 the potential.}
 \label{contours}
\end{figure}

In figure \ref{contours}, contour plots of the potential for
certain values of the parameters $\zeta$ and $\eta$ are shown to
illustrate the qualitative features of the extrema of the
potential.

\section{Summary and Conclusions}

 The model, although in the nature of a toy model has many realistic features. It has a
 solution featuring clash-of-symmetric Higgs kink configurations in a $4+1$ Randall-Sundrum
 space-time. Gravity is localized to the dynamically generated brane,and the RS limit of the solution
is well defined. For the chosen Higgs kink and metric
configurations, the potential has to have a certain sextic form,
whose properties we have studied in some depth.The symmetry
breaking pattern varies as a function of the extra dimension
coordinate $w$, and displays the clash of symmetries phenomenon.
At all points $|w|<\infty$, both $U(1)$ are broken with alternate
$U(1)'s$ restored as $w\rightarrow\pm\infty$. the spontaneous
breaking of the discrete symmetry guarantees topological stability
for the matter-gravity induced brane.

 This work sets the stage for incorporating gravity into more complicated models
displaying the clash of symmetry idea. Our eventual aim is to
construct a realistic brane-world model displaying clash of of
symmetries to induce spontaneous symmetry breaking.

\section*{Acknowledgements}
I would like to thank Dr. Cosmin Macesanu for helpful discussions.
Work supported in part by the U.S.Department of Energy (DOE) Grant
No. DE-FG02-85ER40237.


\begin{thebibliography}{7}

\bibitem{Davidsonet al}
The work reported here is based on the following papers: A.
Davidson, B.F. Toner, R.R.Volkas and K.C. Wali, Phys.Rev
D65,125013; J.S.Rozowsky, R.R.Volkas and K.C.Wali, Physics Letters
B 580 (2004)249-256; G. Dando, A. Davidson, D.P. George,
R.R.Volkas and K.C. Wali, Phys.Rev D, 72, 045016 (2005) For
details and references to related work, please see these papers.
\end{thebibliography}
\end{document}